\documentclass[prd,twocolumn,showpacs,preprintnumbers,nofootinbib,amsmath,amssymb,nobalancelastpage,superscriptaddress,bibliography]{revtex4-1}
\usepackage{graphicx}
\usepackage{bm}
\usepackage{dcolumn}
\usepackage{amsmath}
\usepackage{amssymb}
\usepackage{color}
\usepackage{url}

\usepackage{hyperref}

\usepackage{aas_macros}

\usepackage[normalem]{ulem}

\newcommand{\cm}{{\mathrm{cm}}}
\newcommand{\second}{{\mathrm{s}}}
\newcommand{\TeV}{\mathrm{TeV}}
\newcommand{\GeV}{\mathrm{GeV}}
\newcommand{\MeV}{\mathrm{MeV}}

\newcommand{\sA}{\sigma}
\newcommand{\sigv}{\langle \sA v \rangle}

\newcommand{\degr}{^\circ}

\newcommand{\prob}{\mathrm{P}}

\newcommand{\PSF}{\mathrm{PSF}}

\newcommand{\ch}{\mathrm{ch}}

\newcommand{\bX}{\mathbf{X}}

\newcommand{\nvec}{\mathbf{n}}

\newcommand{\Epeak}{E_\mathrm{peak}}

\begin{document}

\title{Astrophysical explanations of suspected dark matter signals in dwarf galaxies}
\author{Alex Geringer-Sameth}
\email{a.geringer-sameth@imperial.ac.uk}
\affiliation{Department of Physics, Imperial College London, Blackett Laboratory, Prince Consort Road, London SW7 2AZ, UK}
\affiliation{McWilliams Center for Cosmology, Department of Physics, Carnegie Mellon University, Pittsburgh, PA 15213, USA}

\author{Savvas M. Koushiappas}
\email{koushiappas@brown.edu}
\affiliation{Department of Physics, Brown University,  Providence, RI 02912, USA}

\author{Matthew G. Walker}
\email{mgwalker@andrew.cmu.edu}
\affiliation{McWilliams Center for Cosmology, Department of Physics, Carnegie Mellon University, Pittsburgh, PA 15213, USA}

\author{Vincent Bonnivard}
\affiliation{LPSC, Universit\'e Grenoble-Alpes, CNRS/IN2P3, 53 avenue des Martyrs, 38026 Grenoble, France}

\author{C\'eline Combet}
\affiliation{LPSC, Universit\'e Grenoble-Alpes, CNRS/IN2P3, 53 avenue des Martyrs, 38026 Grenoble, France}

\author{David Maurin}
\affiliation{LPSC, Universit\'e Grenoble-Alpes, CNRS/IN2P3, 53 avenue des Martyrs, 38026 Grenoble, France}
\date{\today}

\begin{abstract}

We present methods to assess whether gamma-ray excesses towards Milky Way dwarf galaxies can be attributed to astrophysical sources rather than to dark matter annihilation. As a case study we focus on Reticulum~II, the dwarf which shows the strongest evidence for a gamma-ray signal in Fermi data. Dark matter models and those with curved energy spectra provide good fits to the data, while a simple power law is ruled out at $97.5\%$ confidence. We compare RetII's spectrum to known classes of gamma-ray sources and find a useful representation in terms of spectral curvature and the energy at which the spectral energy distribution peaks. In this space the blazar classes appear segregated from the confidence region occupied by RetII.  Pulsars have similar gamma-ray spectra to RetII but we show that RetII is unlikely to host a pulsar population detectable in gamma rays. Tensions with astrophysical explanations are stronger when analyzing 6.5 years of Pass 7 than with the same amount of Pass 8 data, where the excess is less significant. These methods are applicable to any dwarf galaxy which is a promising dark matter target and shows signs of gamma-ray emission along its line of sight.

\end{abstract}

\pacs{}

\maketitle

Dark matter is the dominant form of mass in the Universe but has, so far, been characterized only through its gravitational effects on astronomical scales. Its microscopic nature, holding fundamental implications for particle physics and cosmology, has yet to be revealed. Astrophysical searches for the high energy particles produced if dark matter annihilates with its antiparticle are a promising way to discover and characterize weakly interacting massive particles with a mass in the GeV--TeV range~\cite[see, e.g.][]{1996PhR...267..195J,2000RPPh...63..793B,2005PhR...405..279B}.

Dark matter annihilation ought to take place anywhere in the universe where dark matter particles encounter each other with sufficient frequency. The annihilation products typically result in the generation of gamma rays, which suffer little deflection or absorption on their way to Earth. This motivates a great variety of searches for anomalous gamma-ray emission in different targets~\citep[e.g.][]{2005PhR...405..279B}, including the Galactic Center~\citep[e.g.][]{2011PhLB..697..412H,2014PhRvD..90b3526A,2015PhRvD..91f3003C}, the Galactic halo~\citep[e.g.][]{2010NuPhB.840..284C,2011PhRvD..83l3516B,2018arXiv180404132C}, Milky Way dwarf galaxies~\citep[e.g.][]{1990Natur.346...39L,2015PhRvD..91h3535G,2015PhRvL.115w1301A}, galaxy groups and clusters~\citep[e.g.][]{2010JCAP...05..025A, 2012JCAP...07..017A,2018PhRvL.120j1101L}, large scale structure~\citep[e.g.][]{2011MNRAS.416.2247X,2014PhRvD..90b3514A,2015PhRvL.114x1301R}, and in the isotropic gamma-ray background~\citep[e.g.][]{2010JCAP...04..014A,2010JCAP...11..041A,2015JCAP...09..008T}.

In searches for dark matter annihilation that gives rise to emission over a continuous energy range (as opposed to a monoenergetic gamma-ray line), conventional astrophysical processes produce gamma rays which compete with the (possibly subdominant) dark matter signal. In this regard, Milky Way dwarf spheroidal galaxies are unique as compared with other targets: they are dark matter dominated systems which contain no known sources of astrophysical emission, making them particularly clean laboratories for dark matter annihilation searches. 

In recent years, dwarf searches have benefited tremendously from full-sky observations of the Large Area Telescope on board the Fermi Gamma-Ray Space Telescope (Fermi LAT)~\citep{2009ApJ...697.1071A}, analysis techniques capable of combining observations of many targets~\citep{2011PhRvL.107x1303G,2011PhRvL.107x1302A,2012APh....37...26M,2013JCAP...03..018S,2014PhRvD..90k2012A,2014PhRvD..89d2001A,2015PhRvD..91h3535G,2015PhRvL.115w1301A,2016JCAP...02..039M,2017PhRvD..95h2001A,2018ApJ...853..154A}, the discovery of new Milky Way satellites~\citep[e.g.][]{2010AdAst2010E..21W,2013NewAR..57..100B,2015PhRvD..91f3515H}, and the characterization of the dark matter distributions within these systems~\citep[e.g.][]{1998ARA&A..36..435M,2007PhRvD..75h3526S,2011ApJ...733L..46W,2011MNRAS.418.1526C,2012CoPhC.183..656C,2015ApJ...801...74G,2015MNRAS.453..849B,2015MNRAS.451.2524M,2016PhRvD..93j3512E,2016MNRAS.461.2914H,2016PhRvD..94f3521S,2017PhRvD..95l3012K,2016JCAP...07..025U,2017MNRAS.466..669C,2018arXiv180206811P}.

With the increasing sensitivity of this effort, it is important to consider how evidence of annihilation in dwarfs might first present itself. While the first goal is always to detect any gamma-ray excess, when indications of one appear we must be prepared to rigorously evaluate whether it originates from dark matter annihilation.  In this paper we consider such ``next steps'' that can be taken to test a dark matter hypothesis when a signal presents itself.

This work is motivated by the 2015 discovery of Reticulum II (RetII), a nearby Milky Way dwarf galaxy found in photometric data from the Dark Energy Survey (DES)~\cite{2015ApJ...805..130K,2015ApJ...807...50B} and confirmed spectroscopically as a system dynamically dominated by dark matter~\cite{2015ApJ...808..108W,2015ApJ...808...95S,2015ApJ...811...62K}. Intriguingly, analysis of 6.5 years of Fermi data reveals a gamma-ray excess between about 2 and 10~GeV significant at the $p=0.0001$ to $0.01$ level, depending on how the background is modeled~\cite{2015PhRvL.115h1101G} ($p$ being the probability that background processes alone can generate such ``signal-like'' data).

While this finding was confirmed by an independent analysis~\cite{2015JCAP...09..016H} of the same data, known as ``Pass 7'', the Fermi-LAT and DES collaborations found a decreased significance, $p=0.05$, for RetII using a reprocessing of the raw LAT data (``Pass 8'') over a similar 6 year baseline~\cite{2015ApJ...809L...4D} (and a  background model analogous to the one that yielded $p=0.01$ above). Subsequent analyses, using $\sim6-7$ years of data, also show decreased significance in Pass 8~\citep{2017ApJ...834..110A,2018ChPhC..42b5102Z} as compared with Pass 7. 

Since these last studies were performed, however, the Pass 8 significance has apparently continued to rise. Recently 
 \citet{2018arXiv180506612L}  presented a 9-year analysis that closely follows the procedure of the 6-year studies~\citep{2015PhRvL.115w1301A,2015ApJ...809L...4D,2017ApJ...834..110A}. They find that RetII steadily grows in significance from three to six to nine years in Pass 8.  They give an uncalibrated significance of $TS=13.5$ ($TS$ being twice the log-likelihood ratio between the dark matter annihilation and background-only hypotheses). While they do not attempt to quantify the trials factor due to testing multiple dark matter masses, annihilation channels, and halo profiles, the local significance (assuming $\chi_1^2/2$ statistics) is $p \sim 10^{-4}$ (compare with $p\sim 10^{-6} - 10^{-5}$ in Pass 7~\citep[Fig.~2 in][]{2015PhRvL.115h1101G}; a trials factor of $\sim3 - 10$ usually suffices for testing multiple masses and channels~\cite{2015PhRvD..91h3535G}). We analyze 6.9 and 10 years of Pass 8 data with the identical method described in~\cite{2015PhRvL.115h1101G} and also observe a rise in significance. The reason for the change in significance between Pass 7 and Pass 8 remains unclear (we discuss consistency in Sec.~\ref{sec:discussion}). Nonetheless, in both data sets RetII possesses the most significant gamma-ray signal of any dwarf galaxy~\cite{2018arXiv180506612L,2015PhRvL.115h1101G,2015PhRvD..91h3535G,2015JCAP...09..016H}.

We take RetII as a case study in the first indication of excess gamma rays from the direction of a dwarf galaxy and consider further hurdles that a dark matter interpretation must overcome. That is, we operate under the assumption that there is a gamma ray source along the direction toward RetII and then seek to characterize this source (we discuss the basis of this assumption in Sec.~\ref{sec:significance}). We emphasize that a dark matter origin cannot be established as long as there is a plausible astrophysical explanation for the gamma-ray excess. The goal of this work is to rigorously address this possibility. 

One strategy is to observe the dwarf at longer wavelengths to try to identify a possible astrophysical counterpart. In a dedicated radio observation, \citet{2017JCAP...07..025R} identified two blazar candidates (BL Lacs) among the sources in RetII's vicinity. As a population, blazars are often associated with gamma-ray emission. However, the distributions of radio, optical, and X-ray fluxes of gamma-ray loud vs. gamma-ray quiet blazars are highly overlapping~\citep[e.g.][]{2015ApJ...810...14A}, making the prediction of any individual blazar's gamma-ray flux quite challenging. At higher frequencies,~\citet{2016A&A...595A..25S} report a detection of 511~keV emission from the direction of RetII, which the authors suggest may be due to an accreting black hole (microquasar) within RetII~\citep{2016A&A...595A..25S,2017arXiv170706871S}. Whether or not this speculative scenario entails gamma ray emission remains to be seen.

In this work we look deeper into the gamma-ray data to assess astrophysical explanations of the excess. Possibilities include a population of gamma-ray emitting objects in the dwarf galaxy itself or a chance alignment with an unrelated, distant gamma-ray source. We begin by quantifying the goodness of fit of dark matter annihilation and astrophysical spectra to the gamma-ray data (Sec.~\ref{sec:GOF}). Then we consider each of the astrophysical possibilities: we compare the shape of RetII's energy spectrum to known classes of gamma-ray sources (Sec.~\ref{sec:3fgl}) and assess whether RetII hosts one or more gamma-ray emitting pulsars (Sec.~\ref{sec:pulsars}). In a companion paper we evaluate a dark matter annihilation hypothesis for the signal.

\section{Data selection}
This work considers both Pass 7 and Pass 8 Fermi-LAT data sets. These are two separate reductions of the raw spacecraft data into reconstructed lists of events along with the associated instrument response functions. To facilitate comparison between them we use data collected during the time when both are available: between Fermi mission weeks 9 and 368 (August 4, 2008 to June 24, 2015; 6.9 years). Pass 7 data is not available after this time. For Pass 7, events and instrument response functions are obtained with version \texttt{v9r33p0} of the Fermi Science Tools\footnote{\url{http://fermi.gsfc.nasa.gov/ssc/}}. We extract Pass 7 Reprocessed SOURCE class events within $15\degr$ of RetII using \texttt{gtselect} with the recommended \texttt{zmax=$100\degr$}. Events must be detected within ``good time intervals'' found using \texttt{gtmktime} with recommended filter \texttt{DATA\_QUAL==1 \&\& LAT\_CONFIG==1} and \texttt{roicut=no}. The instrument response (PSF and exposure) in the direction of RetII are obtained by running \texttt{gtselect} with a radius of $0.5\degr$, \texttt{gtmktime} with the above filter but with \texttt{roicut=yes}, \texttt{gtltcube} with default options, and \texttt{gtpsf} with 17 log-spaced energies between 133.3~MeV and 1.333~TeV, \texttt{thetamax=$10\degr$}, and \texttt{ntheta=500}. Pass 8 SOURCE events and instrument responses are found using the same procedure except we use version \texttt{v10r0p5} of the Science Tools, the recommended \texttt{zmax = $90\degr$} in \texttt{gtselect}, and \texttt{DATA\_QUAL>0 \&\& LAT\_CONFIG==1} in \texttt{gtmktime}. Exposures and PSFs agree with those computed with the latest version \texttt{v11r5p3} of the Science Tools at the subpercent level.

We define a region of interest (ROI) as a circle of radius $0.5\degr$ containing events with energies between 0.5 and 300~GeV. Events in the ROI centered on RetII are used to consider various models for the signal. \citet{2015PhRvL.115h1101G} adopted 1~GeV as the lower end of the energy range. This work includes energies down to 0.5~GeV in order to explore a variety of dark matter and astrophysical models. Dark matter annihilation spectra have shapes which may be a good fit to the data between 1 and 300~GeV but not at energies below 1 GeV. Reducing the energy threshold to 0.5~GeV allows us to better evaluate various interpretations of the data. Due to the energy-dependent point spread function (PSF) of the Fermi LAT, lowering the energy threshold potentially allows for contamination of an ROI by gamma rays from nearby point sources. However, in the case of RetII such sources can be safely ignored\footnote{The nearest source in Fermi's Third Source Catalog (3FGL)~\cite{2015ApJS..218...23A} is a BL Lac blazar, 1RXS~J032521.8-563543, located $2.9\degr$ from RetII. Adopting the spectral model from the 3FGL and the Pass 7 PSF and exposure in the direction of RetII, this source is expected to contribute 0.3 events to the RetII ROI (within uncertainties in the spectral model this can rise to 0.36 events). Other nearby sources (at $3.7\degr$ and $4.4\degr$) contribute significantly fewer events. In contrast, Galactic diffuse and isotropic gamma-ray backgrounds are expected to contribute 140 events to the RetII ROI.   }

\section{Background model and the existence of a source toward RetII}

\subsection{Adopted background model \label{sec:bgdef}}
For the gamma-ray background in the direction of RetII we adopt the Poisson background model used in~\cite{2015PhRvL.115h1101G}: the number of background events in the RetII ROI is a Poisson variable; background events are distributed isotropically within the $0.5\degr$ ROI and their energies are independent samples from a given energy spectrum. We adopt the background energy spectrum derived by the Fermi collaboration\footnote{\url{http://fermi.gsfc.nasa.gov/ssc/data/access/lat/BackgroundModels.html}}. It is the sum of an isotropic component\footnote{Pass 7: \texttt{iso\_source\_v05.txt}, Pass 8: \texttt{iso\_P8R2\_SOURCE\_V6\_v06.txt}} and a diffuse interstellar component\footnote{Pass 7: \texttt{gll\_iem\_v05\_rev1.fit}, Pass 8: \texttt{gll\_iem\_v06.fits}}. The diffuse flux is averaged within a circle of radius $1\degr$ centered on RetII (the effect of changing the size of this region is negligible). As shown in~\cite{2015PhRvL.115h1101G} this model is a very good fit to the average background within $10\degr$ of RetII above 0.2~GeV. We denote the expected background flux $dF_b(E) / dEd\Omega$ (flux of background events per energy per solid angle).

\subsection{Detection significances \label{sec:significance}}
As mentioned in the introduction, two different ways of modeling the gamma-ray background yield two different detection significances for RetII. We reflect on the meaning of this discrepancy and show that comparing the two background models can yield inferences about the presence of a source in the direction of RetII.

Significance represents the degree to which we can reject a ``background-only'' null hypothesis as an explanation of the data. The $p$ value is the probability of obtaining the observed data (or data more ``signal-like'', as quantified by a test statistic) if the null hypothesis were true. For source detection the hypothesis is of the form: The detected events within the RetII ROI are produced by ``background processes'' only. Under the background model of Sec.~\ref{sec:bgdef} the significance of RetII using Pass 7 data is $p \approx 10^{-4}$ (see~\citep{2015PhRvL.115h1101G} and Table~\ref{tab:gof78} of this paper).

A second background model and resulting significance test is provided by the ``empirical background'' technique~\citep{2011PhRvL.107x1303G,2015PhRvD..91h3535G} (see also~\citep{2014PhRvD..89d2001A,2015PhRvD..91f1302C,2015PhRvL.115w1301A,2017ApJ...834..110A} for an analogous ``blank sky locations'' method). Here, the same test statistic applied to the RetII ROI is applied to random locations within $10\degr$ of RetII, building up the empirical probability distribution of the test statistic under the background hypothesis. The underlying assumption is that whatever background processes are at work near RetII are also at work in the direction of RetII. Then the $p$ value is the fraction of all sampled background ROIs that look more ``signal-like'' than the RetII ROI (i.e. have a larger value of the test statistic). \citet{2015PhRvL.115h1101G} find $p \approx 0.01$ from the Pass 7 data using this method.

If we sharpen up what is meant by ``background processes'' we will see that the two different significances for RetII come from testing two distinct hypotheses.

The first tested hypothesis is that the background model of Sec.~\ref{sec:bgdef} with no additional contributions can explain the RetII data. The energy spectrum $dF_b(E) / dEd\Omega$ is based mainly on physical models of the Milky Way's interstellar medium~\citep{2016ApJS..223...26A} (e.g. cosmic ray interactions with gas), diffuse structures like the Fermi bubbles,  and isotropic emission over the whole sky (which also accounts for cosmic ray contamination). Such a fundamentally diffuse process of emission is governed by Poisson statistics, with an energy spectrum changing smoothly from place to place. In particular, the model does not include discrete ``bright'' point sources, i.e. those with fluxes at or above the level of the diffuse processes.

Obtaining $p\approx10^{-4}$ for this hypothesis means that the emission towards RetII is either due to a rare statistical fluctuation or to the adoption of an incorrect spectral model (this is the usual issue of statistics vs. systematics). If the spectral model is incorrect it can be for two reasons: the presence of a localized ``bright'' source of gamma rays or the inadequate modeling of the diffuse physical processes.

The last explanation (no additional source but a mismodeling of the diffuse processes) is unlikely for a few reasons. First, the background model is a very good fit overall to the $10\degr$ region surrounding RetII, showing no systematic deviation from the data~\citep{2015PhRvL.115h1101G,2015ApJ...809L...4D}. Second, the background spectrum has a significantly different shape from the observed RetII spectrum~\citep[Fig.~1 of][]{2015PhRvL.115h1101G}, and a good fit cannot be obtained by changing its normalization (e.g. if the amount of gas along the line of sight were underestimated in the model; see also~\citep{Hoof2018}). In fact, there is no place in the sky where the Fermi diffuse model has the shape of the RetII spectrum\footnote{We determined this by extracting the spectrum of the Fermi Pass 7 diffuse interstellar emission model within every $0.125\degr \times 0.125\degr$ pixel covering the whole sky and adding it to the isotropic spectral model. At no location does the spectral energy distribution $E^2 dF/dE$ peak above 1~GeV (compare with $\gtrsim 2~\GeV$ for RetII; see Secs.~\ref{sec:GOF} and~\ref{sec:3fgl}), with 99.8\% of the sky having $E^2 dF/dE$ peak below 0.7~GeV (the Pass 8 model gives similar results).}, which we take as an indication that conventional diffuse processes cannot give rise to such a spectrum. Third, the RetII excess is localized. If the diffuse model were incorrect there would likely be a highly spatially-correlated excess. Furthermore, the diffuse model shows no large variations or complicated behavior near the location of RetII, which is $50\degr$ off the galactic plane. Finally, the RetII signal is undiminished if, instead of SOURCE events, we compute the significance using Pass 7 ULTRACLEAN events, which are a subset of SOURCE events reconstructed with higher quality and suffering a smaller cosmic ray contamination.

With cautious confidence in the diffuse model, the remaining possibilities are either a ``bright'' source toward RetII or a Poisson fluctuation in the detected events. We can use the results of the hypothesis test based on the empirical background model to explore this further.

Whereas $p\approx 10^{-4}$ is the probability that diffuse processes such as cosmic ray interactions with gas and extragalactic isotropic emission can explain the RetII data, $p\approx 0.01$ is the probability of an excess from {\em all mechanisms} besides emission from a dwarf galaxy (i.e. from any cause other than dark matter annihilation within the dwarf). The additional mechanisms in this more inclusive hypothesis are those stated above: the presence of additional gamma-ray sources along the line of sight and systematic deviations from the assumed diffuse model. That this probability of 0.01 is relatively high by particle physics standards means that, for this set of data {\it and} using this particular test statistic, we cannot reject the hypothesis that there is no dark matter annihilation taking place in RetII. However, as we show next, we can use this information in a back of the envelope calculation that shows that the RetII data are much more likely to be due to a source than to a background fluctuation.

Consider all the significant-looking sky locations like RetII's.  How many contain ``bright'' sources (i.e. those with flux above the diffuse level) and how many are Poisson fluctuations of the diffuse processes (assuming they are correctly modeled)? Let $\prob(F_b)$ be the probability that the brightest source in a random ROI has a flux below the diffuse background level $F_b$. We seek the probability that a given ROI with ``RetII-like'' gamma-ray data $D$\footnote{More precisely, $D$ is the statement that the test statistic for the ROI is larger than the one observed in RetII.} does not contain a bright source: $\prob(F_b \mid D)$. This can be rewritten in terms of $p$ values: $\prob(F_b \mid D) = \prob(D \mid F_b) \prob(F_b) / \prob(D)$, where the two hypothesis tests discussed correspond to $\prob(D \mid F_b) \approx 10^{-4}$ and $\prob(D)\approx 0.01$. Since $\prob(F_b) < 1$ we must have that $\prob(F_b \mid D) < 0.01$\footnote{Applying this argument to the actual RetII ROI would take us into Bayesian territory. In that case $P(D)$ is greater than 0.01, but to quantify it we would have to assign degrees of belief to the various particle properties of dark matter and the parameters describing RetII's dark matter halo.}. In words, the probability of a statistical fluctuation is less than 1\% and the probability that there is a source with flux above the diffuse background level is greater than 99\%. Given the data, and based only on the observable statistics of the gamma-ray sky, it is over 100 times more likely that the RetII excess arises from an above-background source rather than from a Poisson fluctuation of the diffuse background.

This rough calculation has no bearing on whether or not the emission is caused by dark matter annihilation. Rather, it justifies us taking the simple existence of a source as the {\em starting point} for our explorations here. We note that commonly used test statistics~\citep[e.g.][]{1996ApJ...461..396M,2015PhRvD..91h3535G,2015PhRvL.115w1301A} are designed to be powerful at rejecting the diffuse model as the null hypothesis, not at distinguishing a dark matter signal from a previously unidentified astrophysical point source. This paper focuses on this second question (see also~\citep{2015JCAP...09..016H,2015PhRvD..91f1302C,2015JCAP...05..056L,2017AJ....153..253M} for progress in incorporating unknown source populations at the level of the likelihood).

\section{Gamma-ray likelihood}

The gamma-ray observable $\bX_\gamma$ is the list of events $i$ with energies $E_i$ and angular separations $\phi_i$ from the center of the ROI. Dividing the events into bins of energy and angular separation we have $n_j$ events in bin~$j$. Model parameters (e.g. dark matter particle properties, energy spectrum shape parameters, or those describing RetII's dark matter halo) are denoted by $\theta$. The expected number of counts in bin $j$ is $\mu_j(\theta)$. For the adopted Poisson background model, the probability (or likelihood) of observing the set $\nvec \equiv (n_1, n_2, \dots)$ given model parameters $\theta$ is simply the product of Poisson distributions:
\begin{equation}
\prob(\nvec \mid \theta) = \exp \left( - \sum_j \mu_j(\theta) \right) \prod_j \frac{\mu_j(\theta)^{n_j}}{n_j !}.
\label{eqn:likelihoodbinned}
\end{equation}

The expected counts can be divided into source (signal) and background components: $\mu_j(\theta) = B_j + S_j(\theta)$, where $B_j$ and $S_j(\theta)$ are integrals of the differential expected counts $b(E)$ and $s(E,\phi \mid \theta)$ over the $E$ and $\phi$ range of bin $j$.

The differential background $b(E)$ (predicted background events per energy per solid angle) is 
\begin{equation}
b(E) = \frac{d F_b(E)}{dE d\Omega} \epsilon(E),
\end{equation}
where $dF_b(E) / dEd\Omega$ is the adopted background flux model of Sec.~\ref{sec:bgdef} and $\epsilon(E)$ is the Fermi-LAT exposure (effective area $\times$ time) in the direction of RetII.

The differential signal $s(E,\phi \mid \theta)$ for a point source depends on energy and angular separation from RetII and on the model parameters $\theta$:
\begin{equation}
s(E,\phi \mid \theta) = \frac{d F(E \mid \theta)}{dE} \epsilon(E) \PSF(\phi \mid E),
\label{eqn:signalfluxpoint}
\end{equation}
where $dF(E \mid \theta)/dE$ is the source photon flux per energy and the $\phi$ dependence is governed entirely by the instrument's PSF\footnote{In this work we do not model the finite energy resolution of the LAT ($\Delta E/E \lesssim 0.1$ for $E \gtrsim 0.5\,\GeV$) as the spectra we consider are much broader than this.}. We discuss the choice to model RetII as a point source rather than an extended one in Sec.~\ref{sec:DMdef}.

Our statistical tests will be based on an unbinned likelihood. As the size of the bins in $E$ and $\phi$ shrink to zero Eq.~\ref{eqn:likelihoodbinned} becomes
\begin{equation}
\prob(\bX_\gamma \mid \theta) \propto \exp\left( -\int (s+b) dEd\Omega \right) \prod\limits_i (s_i + b_i),
\label{eqn:likelihoodunbinned}
\end{equation}
where the integral is over the entire ROI (i.e. all energies and angular separations). The product in Eq.~\ref{eqn:likelihoodunbinned} is over the individual observed events, i.e. $s_i = s(E_i, \phi_i \mid \theta)$.

In the limit of small bins the constant of proportionality in Eq.~\ref{eqn:likelihoodunbinned} goes to zero. It is convenient to normalize the probability by a term which does not depend on the model parameters $\theta$. A likelihood ratio where the denominator is the probability under the background-only model is a convenient choice. Dividing Eq.~\ref{eqn:likelihoodbinned} by itself but with all $S_j=0$ yields a finite limit as the bins become infinitesimal (cf.~Eq.~22 of~\cite{2015PhRvD..91h3535G}):
\begin{equation}
\frac{\prob(\bX_\gamma \mid \theta)}{\prob(\bX_\gamma \mid s=0)} = \exp\left( -\int s\, dEd\Omega \right) \prod\limits_i \left(1 + \frac{s_i}{b_i} \right).
\label{eqn:likelihoodratio}
\end{equation}

\section{Source models \label{sec:sourcemodels}}

We consider two classes of models to describe the gamma ray source toward RetII: phenomenological descriptions of astrophysical sources and dark matter annihilation within RetII.

\subsection{Astrophysical source models}

We model astrophysical sources as point sources with either power law or curved ``log parabola'' spectra. These two functional forms are used to describe the vast majority of gamma-ray sources in the Fermi Third Source Catalog (3FGL)~\cite{2015ApJS..218...23A}. In the 3FGL each source (unless it is a pulsar) is fit with both a power law and a log parabola spectrum. If the log parabola spectrum is found to be a significantly better fit (difference in test statistic greater than 16) it is adopted as the ``spectral type'' in the catalog. Of the 3034 sources in the catalog, 2523 are described by a power law spectrum and 395 are assigned log parabola spectra. The remaining 116 sources are pulsars (and the extremely bright blazar 3C~454.3) and are fit with power laws with exponential or superexponential cutoffs. We consider a pulsar interpretation of the RetII signal in Sec.~\ref{sec:pulsars}. We note that other spectral shapes (e.g. broken power law) may provide better fits to some sources. However, for the purpose of comparing RetII's spectrum to those of known gamma ray sources we adopt the same spectral models used in the 3FGL.

The power law spectrum has two model parameters, a normalization $F_0$ and a slope $\alpha$,
\begin{equation}
\frac{d F(E \mid \theta)}{dE} = F_0 \left( \frac{E}{E_0}\right)^{-\alpha},
\label{eqn:powerlawdef}
\end{equation}
where $E_0$ is an arbitrary reference energy that we fix to 1~GeV.

The log parabola spectrum has an additional curvature parameter $\beta$:
\begin{equation}
\frac{d F(E \mid \theta)}{dE} = F_0 \left( \frac{E}{E_0}\right)^{-\alpha - \beta \log(E/E_0)},
\label{eqn:logparaboladef}
\end{equation}
where $\log$ is the natural logarithm. In the 3FGL the reference energy, called the pivot energy $E_p$, varies from source to source. Changing the reference energy changes the parameter $\alpha$~\cite{2012ApJS..199...31N}: ${\alpha(E_p) = \alpha(E_0) + 2\beta \log(E_p/E_0)}$. We convert the $\alpha(E_p)$'s given in the 3FGL to ${\alpha(E_0=1\,\GeV)}$ for this work.

\subsection{Dark matter annihilation \label{sec:DMdef}}

For dark matter annihilation the model parameters are  $\theta = (M, \sigv, \ch, J)$, with the first three representing the dark matter particle mass, its velocity-averaged annihilation cross section, and the annihilation channel (i.e. Standard Model final state). We treat RetII as a point source of gamma-rays (see below). Therefore, as far as gamma-ray emission is concerned, its dark matter halo is parameterized by a single quantity $J$, the integral over the halo volume of the dark matter density squared divided by the line-of-sight distance squared. The dark matter annihilation flux to be used in Eq.~\ref{eqn:signalfluxpoint} is given by (e.g.~\cite{2015PhRvD..91h3535G})
\begin{equation}
\frac{d F(E \mid\theta)}{dE} = \frac{\sigv J}{8\pi M^2} \frac{dN_\gamma(E)}{dE},
\label{eqn:signalfluxDM}
\end{equation}
where $dN_\gamma/dE$ is the number of gamma-rays emitted per annihilation (per energy) for the given final state channel and mass $M$. For $dN_\gamma/dE$ we adopt the spectra computed by \citet{2011JCAP...03..051C}, which include electroweak corrections~\cite{2011JCAP...03..019C}. For point-source emission $J$ is exactly degenerate with $\sigv$ in Eq.~\ref{eqn:signalfluxDM} and we treat $\sigv J$ as a single parameter which normalizes the amplitude of the signal. In this way our results are independent of any particular choice of $J$.

Equations~\ref{eqn:signalfluxpoint} and~\ref{eqn:signalfluxDM} represent a point source approximation. To be accurate, $J$ must be replaced with the $J$-profile $dJ(\phi)/d\Omega$ (e.g.~\cite{2015ApJ...801...74G}) and convolved with the PSF as described in~\cite{2015PhRvD..91h3535G}. The $J$-profile is the integral of the square of the dark matter density along the line of sight as a function of the angle $\phi$ from the center of the dwarf. The use of Eqs.~\ref{eqn:signalfluxpoint} and~\ref{eqn:signalfluxDM} is justified if the $J$-profile is much narrower than Fermi's PSF. The 68\% containment angle of the gamma-ray PSF for the RetII observation is about $0.5\degr$ at 2~GeV and decreases to $0.2\degr$ at 10~GeV. Interestingly, the median posterior estimate of the 68\% containment angle for RetII's $J$-profile, as measured by \citet{2015ApJ...808L..36B}, is roughly $1\degr$, with half the sampled $J$-profiles being more extended. However, $0.5\degr$ corresponds to 260~pc at the distance to RetII (30~kpc), while the halflight radius of RetII is only 58~pc~\cite{2018arXiv180408627M} and the outermost spectroscopically confirmed member star is at a projected distance of 90~pc~\cite{2015ApJ...808..108W}. This means that inferences about the extent of RetII's dark matter halo strongly depend on assumptions about the halo beyond the radius probed by observations. Thus at the present time no firm conclusions can be made about RetII's dark matter distribution on angular scales of $0.5\degr$. We note that~\citet{2015JCAP...09..016H} find that the Pass 7 data do not prefer a departure from the point source assumption for RetII, while \citet{2018arXiv180506612L} find a a slight ($\Delta \chi^2 = 1.3$) preference for extension in 9 years of Pass 8.

\section{Goodness of fit of various spectral models \label{sec:GOF}}

In this section we consider which spectral models are good fits to the gamma-ray data from RetII and which cannot explain the emission.

\subsection{Method \label{sec:GOFmethod}}
We use a likelihood ratio to assess the goodness of fit of the various emission models to the RetII data. Under the null hypothesis we wish to test, the emission is governed by a particular spectral model and associated parameters $\theta_0$. For example, $\theta_0$ might be dark matter annihilation with a given mass, channel, and value of $\sigv J$. Or it could be a log parabola model with a specified $F_0$, $\alpha$, and $\beta$.

A powerful test of the null hypothesis is performed by comparing $\theta_0$ to plausible alternatives using a likelihood ratio. We take these alternatives to be any of the spectral models described in Sec.~\ref{sec:sourcemodels}. The test statistic, a function of the gamma-ray data $\bX_\gamma$, is (e.g.~\cite{Kendall5th,casella2002statistical})
\begin{equation}
\lambda(\bX_\gamma)  = 2 \log \frac{\prob(\bX_\gamma \mid \hat{\theta})}{\prob(\bX_\gamma \mid \theta_0)}.
\label{eqn:likelihoodratiotest}
\end{equation}
In this equation, $\hat{\theta}$ is the model which maximizes the likelihood for the given data set $\bX_\gamma$. The maximization is performed over all spectral types (dark matter, power law, or log parabola) and all parameters within those types (e.g. mass, channel, $F_0$, $\alpha$, $\beta$, etc.). Large values of $\lambda(\bX_\gamma)$ indicate that the hypothesis $\theta_0$ is a poor fit to the data $\bX_\gamma$. We use Eq.~\ref{eqn:likelihoodratio} to compute the likelihood ratio.

Our set of alternative models are not nested, the true values of $\theta$ may lie beyond the boundaries of the parameter space for the log parabola model (see below), and it is unclear whether the number of events in the $0.5\degr$ ROI is large enough to apply Wilks theorem~\citep{wilks1938,Kendall5th}. We therefore simulate large numbers of fake ROIs to directly construct the sampling distribution of $\lambda(\bX_\gamma)$ under a given hypothesis $\theta_0$ (see~\citep{2016MNRAS.458L..84A} for a possible alternative). Background events are generated using the model of Sec.~\ref{sec:bgdef} and signal events using the models of Sec.~\ref{sec:sourcemodels}. The goodness of fit $p$ value for $\theta_0$ is the fraction of realizations with a larger value of $\lambda(\bX_\gamma)$ than obtained for the observed RetII data. We find that, for the best fitting models $\theta_0$, the distribution of $\lambda(\bX_\gamma)$ is not well described by a $\chi^2$ distribution. However, for the 13 best fitting models (11 dark matter, power law, and log parabola) the PDF of $\lambda(\bX_\gamma)$ is fairly well described by a gamma distribution with shape parameter $k\approx3$ and scale parameter $\theta \approx 1.4$, suggesting that a scaled version of $\lambda(\bX_\gamma)$ may be distributed as a $\chi^2$ variable with 6 degrees of freedom\footnote{This holds when the true parameters are sufficiently far from the boundary of parameter space. In the case where the null hypothesis is background-only (i.e. $F_0=0$) about 10\% the samples have $\lambda(\bX_\gamma)=0$ while the rest are gamma-distributed with $k \approx0.8$ and $\theta \approx 2.3$.}.

The maximization of the likelihood is performed over a grid of model parameters (except for the normalizations $\sigv J$ and $F_0$ which can vary continuously). We have chosen the grid to be fine enough so that the results are not sensitive to the discreteness. For the dark matter models the allowed masses run from the mass of the final state particle up to 1~TeV in log-spaced steps where neighboring masses differ by 2\%. For the power law spectrum we consider indices $\alpha$ running from 1 to 3 in steps of 0.01 (the range for 3FGL sources is between 1.1 to 5.7). For the log parabola models, $\alpha$ runs from $-1$ to 5 in steps of 0.05 (the range in the 3FGL is -0.54 to 4.6), and $\beta$ runs from 0.05 to 1 in steps of 0.05 (the range in the 3FGL is 0.03 to 1). Though values of $\beta$ greater than 1 are physical, the upper limit of 1 is imposed in order to more easily compare with the 3FGL, where 1 is the maximum allowable $\beta$. Additionally, the log parabola model is meant to model astrophysical sources and it is appropriate to restrict its parameter space to where such sources are expected to lie. We discuss the relaxation of the $\beta <1$ requirement in Sec.~\ref{sec:3fgl}.

\subsection{Results}

We find the best fitting model parameters (those which maximize the likelihood in Eq.~\ref{eqn:likelihoodratio}) for each spectral class: power law, log parabola, and dark matter for each annihilation channel. We then test whether each of these best fit models is actually a good fit to the RetII data using the likelihood ratio test described above. That is, we set $\theta_0$ to a best fit model, generate fake data $\bX_\gamma$ under this model to find the distribution of $\lambda(\bX_\gamma)$, and find what fraction of fake data sets have $\lambda(\bX_\gamma)$ higher than that observed for RetII. The resulting $p$ values are shown in Table~\ref{tab:gof78}.

\begin{table*}[]
\caption{\label{tab:gof78}Goodness of fit for the best fitting models to the gamma-ray signal from RetII. The first two rows are for log parabola (Eq.~\ref{eqn:logparaboladef}) and power law spectra (Eq.~\ref{eqn:powerlawdef}). The third row gives the goodness of fit for the background-only model (i.e. $F_0=0$). The last 11 rows correspond to gamma-ray spectra of dark matter annihilating into various Standard Model final states (Eq.~\ref{eqn:signalfluxDM}). The first three columns give the best fitting model parameters, while the fourth column is the likelihood ratio of the model compared to the best fitting model (see Eq.~\ref{eqn:likelihoodratiotest}). The last column is the goodness of fit $p$ value of the model: the probability of obtaining a larger $\lambda(\bX_\gamma)$ than observed for RetII if the model were true. Each pair of columns shows results for Pass 7 on the left and Pass 8 on the right.}
\begin{ruledtabular}
\begin{tabular}{dddddddddd}

\multicolumn{2}{c}{$\alpha$} &
\multicolumn{2}{c}{$\beta$}   &
\multicolumn{2}{c}{$F_0$}     &
\multicolumn{2}{c}{$\lambda(\bX_\gamma)$} & 
\multicolumn{2}{c}{Goodness of fit }\\
\multicolumn{2}{c}{} &
\multicolumn{2}{c}{} &
\multicolumn{2}{c}{[$10^{-11} \cm^{-2} \second^{-1} \GeV^{-1}$]} &
\multicolumn{2}{c}{} &
\multicolumn{2}{c}{$p$ value} \\

\multicolumn{1}{c}{$\scriptstyle{\mathrm{Pass~7}}$} & \multicolumn{1}{c}{$\scriptstyle{\mathrm{Pass~8}}$} & \multicolumn{1}{c}{$\scriptstyle{\mathrm{Pass~7}}$} & \multicolumn{1}{c}{$\scriptstyle{\mathrm{Pass~8}}$} & \multicolumn{1}{c}{$\scriptstyle{\mathrm{Pass~7}}$} & \multicolumn{1}{c}{$\scriptstyle{\mathrm{Pass~8}}$} & \multicolumn{1}{c}{$\scriptstyle{\mathrm{Pass~7}}$} & \multicolumn{1}{c}{$\scriptstyle{\mathrm{Pass~8}}$} & \multicolumn{1}{c}{$\scriptstyle{\mathrm{Pass~7}}$} & \multicolumn{1}{c}{$\scriptstyle{\mathrm{Pass~8}}$} \\

\colrule
-0.70 & -1.00^*\footnotemark[1] & 1.00^* & 0.95 &  3.0 &0.80  & 2.5 & 1.3 & 0.73 &0.90\\
2.09 &1.99&             &                                               & 10.6 &  2.8   & 9.9 & 4.9 & 0.025 &0.16\\
\multicolumn{6}{l}{Background-only model}  & 19.9 & 7.0 & \multicolumn{1}{c}{$8.8\times 10^{-5}$} & 0.027\\
\\

\multicolumn{2}{c}{Channel} &
\multicolumn{2}{c}{$M$}   &
\multicolumn{2}{c}{$\sigv_{-26} J_{19.6}\footnotemark[2]$}    \\
\multicolumn{2}{c}{}&
\multicolumn{2}{c}{[GeV]} &
\multicolumn{2}{c}{}\\
\cline{1-6}
\multicolumn{2}{c}{$e^+e^-$}            & 6.2 & 8.1 & 2.9 & 1.7 & 0.0 & 0.0 & 1.0 & 1.0\\
\multicolumn{2}{c}{$\mu^+\mu^-$}    & 6.2 & 8.1 & 6.5 & 3.5 & 0.8 & 0.46 & 0.93 & 0.97\\
\multicolumn{2}{c}{$\tau^+\tau^-$}    & 13.9 & 19.5 & 1.1 & 0.63 & 2.7 & 1.5 & 0.70 & 0.89\\
\multicolumn{2}{c}{$b\bar{b}$}          & 77.6 & 120.5 & 3.8 & 2.2 & 5.1 & 3.0 & 0.29 & 0.51\\
\multicolumn{2}{c}{$q\bar{q}$}          & 34.8 & 55.1 & 1.7 & 0.96 & 5.4 & 3.2 & 0.24 & 0.44\\
\multicolumn{2}{c}{$c\bar{c}$}          & 31.5 & 47.9 & 1.5 & 0.81 & 5.4 & 3.2 & 0.24 & 0.44\\
\multicolumn{2}{c}{$gg$}                    & 32.8 & 48.9 & 1.5 & 0.83 & 5.4 & 3.2 & 0.24 & 0.44\\
\multicolumn{2}{c}{$t\bar{t}$}           & 180.0^* & 195.0 & 11.1 & 4.4 & 5.5 & 3.0 & 0.23 & 0.48\\
\multicolumn{2}{c}{$hh$}                  & 135.3 & 214.3 & 7.8 & 4.6 & 5.6 & 3.2 & 0.23 & 0.45\\
\multicolumn{2}{c}{$W^+W^-$}        & 90.0^* & 99.5 & 6.3 & 2.5 & 5.7 & 3.2 & 0.21 & 0.43\\
\multicolumn{2}{c}{$ZZ$}                  & 100.0^* & 124.6 & 7.0 & 3.2 & 5.7 & 3.3 & 0.21 & 0.42\\

\end{tabular}
\end{ruledtabular}
\footnotetext[1]{An asterisk indicates that the parameter value is at the boundary of its allowed range.}
\footnotetext[2]{$\sigv_{-26} J_{19.6} = (\sigv / 10^{-26} \cm^3 \second^{-1}) (J / 10^{19.6}\, \GeV^2 \cm^{-5})$ and $10^{19.6}\, \GeV^2 \cm^{-5}$ is the median posterior estimate of RetII's $J$-profile integrated within $0.5\degr$~\cite{2015ApJ...808L..36B}.}
\end{table*}

The best fits (highest $p$ values) are for a source with a log parabola spectrum or dark matter particles annihilating into leptons, followed by annihilation into quarks and gauge bosons. For every dark matter model there is at least one particle mass for which the fit is acceptable.

Power law models, on the other hand, are in tension with the Pass 7 RetII data with $p =  0.025$. Specifically, if there were a power law source in the direction of RetII with spectral index of $\alpha = 2.09$ there is only a 2.5\% chance of finding $\lambda(\bX_\gamma)$ as large is it is measured to be. In other words, RetII appears to have a significantly curved spectrum.

Note that Table~\ref{tab:gof78} only shows best fitting models, not uncertainties in model parameters. In the next section we discuss constraints on the log parabola model parameters and a companion paper will explore dark matter parameter space (see also Fig.~4 in \cite{2015PhRvL.115h1101G}).

\begin{figure}
\includegraphics{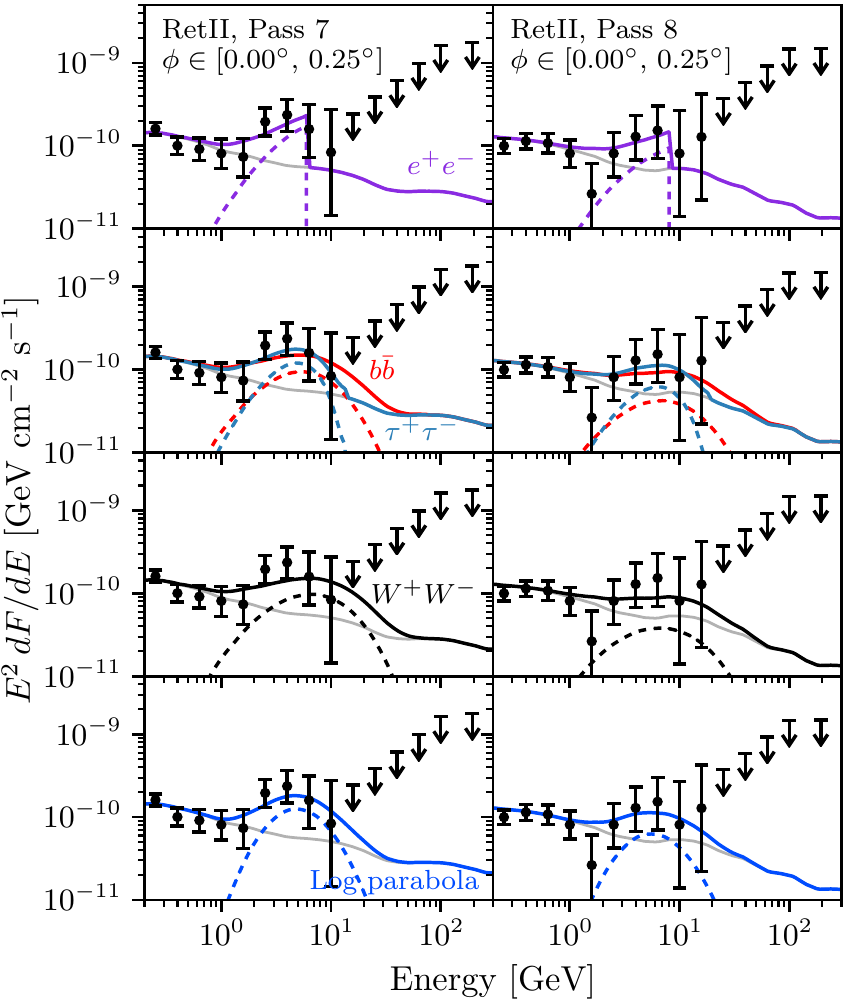}
\caption{\label{fig:speckstackmodels} Energy spectrum of RetII compared with best fitting models. Data points (same for each row) are the observed spectrum derived from events detected within $0.25\degr$ of RetII along with 68\% Poisson error bars. The thin gray curve is the predicted background and dashed curves show the source contribution. Solid colored curves are the sum of background and source. The first three rows are models of dark matter annihilation while the last describes a generic astrophysical source with a curved spectrum. The two columns show results for Pass 7 and Pass 8 data.}
\end{figure}

Figure~\ref{fig:speckstackmodels} shows the best fitting model spectra and compares them with the observed RetII data. The data points show the empirical spectrum of RetII derived from the observed event counts within $0.25\degr$ of RetII. These are binned in energy (5 bins per decade starting at 0.2~GeV) and error bars show 68\% Poisson confidence intervals. The empirical flux is the number of counts divided by the exposure and energy bin width. Model spectra as well as the background spectrum are plotted as curves. Pass 7 and Pass 8 results are shown with their respective best fitting models (we discuss consistency in Sec.~\ref{sec:discussion}).

\begin{figure}
\includegraphics{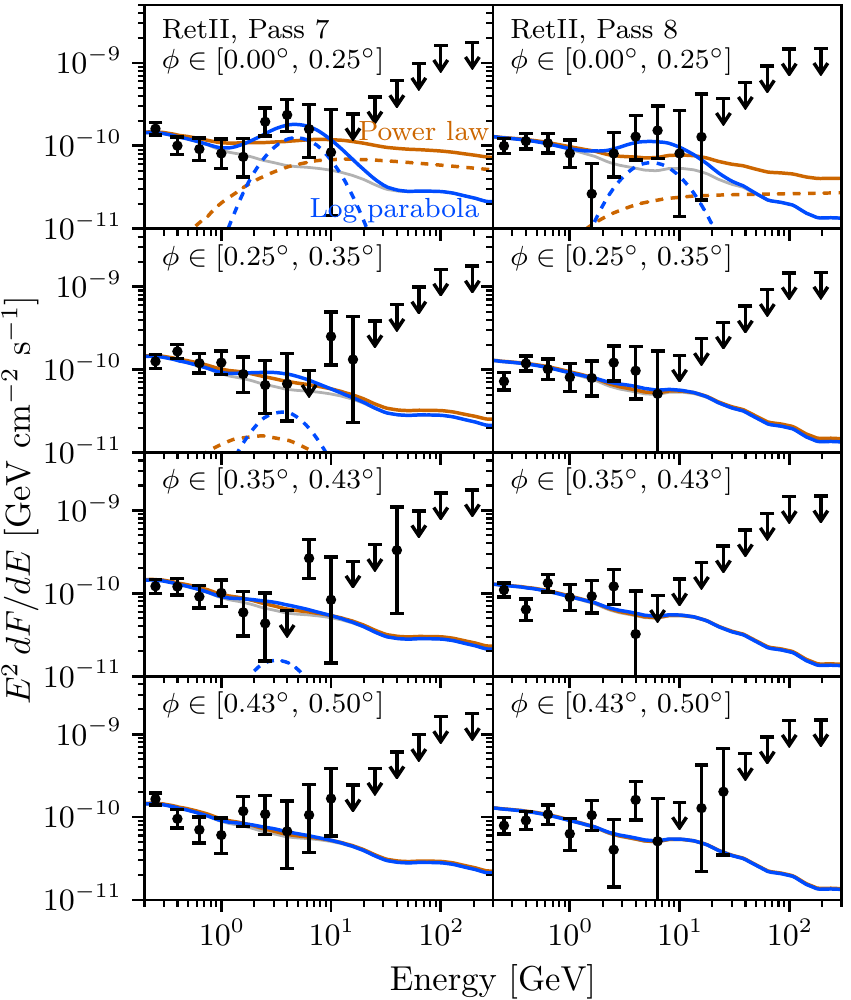}
\caption{\label{fig:speckstackannuli} Same as Fig.~\ref{fig:speckstackmodels} but showing the fits at varying angular separation from RetII. Different rows correspond to spectra constructed from different annuli. The best fit power law and log parabola source models are compared.}
\end{figure}

Figure~\ref{fig:speckstackannuli} illustrates the fit as a function of energy and angular separation from RetII. The $0.5\degr$ RetII ROI is divided into four annuli with equal solid angle, which are shown as different rows. The best fitting log parabola and power law models are plotted. In Figs.~\ref{fig:speckstackmodels} and~\ref{fig:speckstackannuli} the model spectra (Eqs.~\ref{eqn:powerlawdef},~\ref{eqn:logparaboladef}, and~\ref{eqn:signalfluxDM}) are scaled by the PSF integrated within the corresponding annulus in order to compare with data. The energy dependence of the PSF explains why, for example, the power law spectrum is not a straight line.

\section{Comparison with astrophysical populations}

\subsection{\label{sec:3fgl}Sources in the 3FGL}

Among the models meant to describe astrophysical sources, the log parabola spectrum is a perfectly acceptable fit to the data ($p=0.73$ in Pass 7). This motivates a comparison with the various source populations present in the 3FGL. Another likelihood ratio test statistic is used to place constraints on the $\alpha$ and $\beta$ parameters of the log parabola spectrum that can describe the RetII emission. The space of alternative hypotheses (the numerator in Eq.~\ref{eqn:likelihoodratiotest}) is restricted to include only log parabola spectra with $-1 \leq \alpha \leq 5$, $0 \leq \beta \leq 1$, and any value for $F_0$.  To test whether a given set $\alpha, \beta$ is an acceptable fit to the data we maximize the likelihood over the normalization $F_0$ while holding $\alpha$ and $\beta$ fixed. This constrained maximum likelihood value is used as the null hypotheses in the denominator of Eq.~\ref{eqn:likelihoodratiotest}. With fixed numbers of degrees of freedom in the null and alternative hypotheses, we cautiously make use of the $\chi^2$ approximation to the likelihood ratio~\citep[e.g.][]{Kendall5th}. In this case $\lambda(\bX_\gamma)$ should be distributed as $\chi^2$ with 2 degrees of freedom and regions of $\alpha, \beta$ space where $\lambda(\bX_\gamma) > 2.3\, (6.2)$ are ruled out at 68.3\% (95.4\%) significance.

\begin{figure*}
\includegraphics{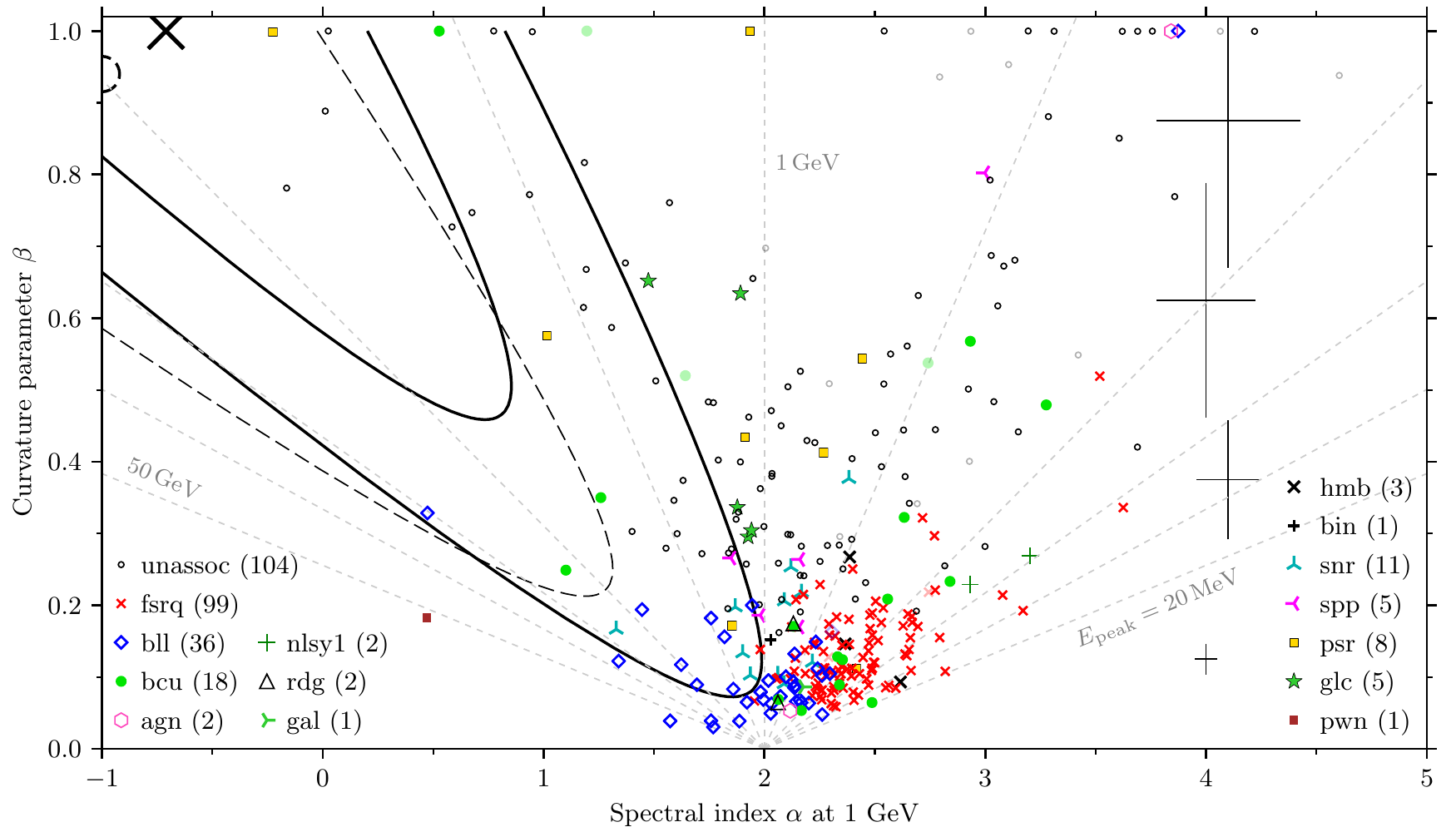}
\caption{\label{fig:logparabola3fgl} The energy spectrum of Reticulum~II compared with sources in the 3FGL. Axes correspond to the parameters of the log parabola spectrum (Eq.~\ref{eqn:logparaboladef}). The large black cross shows the best fit values using Pass 7 data and the solid black contours are 68\% and 95\% confidence intervals on RetII's $\alpha$ and $\beta$ (constrained to the range $-1 < \alpha < 5$ and $0.05 < \beta < 1$). Analogous Pass 8 results shown by the large dashed circle, with the dashed contour representing the 68\% confidence region (the 95\% Pass 8 contour includes the entire plane). Individual points represent sources in the 3FGL, with different marker styles corresponding to different source classes. These include unassociated sources (empty circles), flat spectrum radio quasars (fsrq, red crosses), BL Lacs (bll, empty blue diamonds), and blazars of uncertain type (bcu, filled green circles). Extragalactic sources are listed in the legend on the left, galactic sources on the right (see Table 6 of~\cite{2015ApJS..218...23A} for a complete description of the source classes). The four black error bars running up the right-hand side show typical error bars for the 3FGL points (error sizes are highly correlated with $\beta$, see text). Faded markers indicate 17 sources for which the 3FGL analysis has been flagged as potentially unreliable. Lines of constant $\Epeak$, the energy at which $E^2 dF/dE$ peaks, are shown as gray dashed lines radiating from $(\alpha, \beta)=(2,0)$. In counterclockwise order, the lines correspond to $\Epeak=0.02, 0.05, 0.1, 0.2, 0.5, 1, 2, 5, 10, 20$, and $50~\GeV$.}
\end{figure*}

The contours in Fig.~\ref{fig:logparabola3fgl} show the resulting confidence intervals. The large black cross shows the best fitting parameters for the Pass 7 RetII data, occurring at the edge of the allowable parameter space at $(\alpha, \beta)=(-0.7,1)$. Solid lines show the 68\% and 95\% confidence regions for Pass 7 (the large black dashed circle and dashed black contour show the best fit and 68\% region for Pass 8; the 95\% contour includes the entire figure since $\lambda(\bX_\gamma)<6.2$ for the log parabola model in Pass 8). 

To check the coverage of the confidence intervals we simulated $10^4$ fake data sets for each of 20 points along the 68\% and 95\% contours (using the best fit $F_0$ at that $(\alpha,\beta)$ value). For each fake data set we find the sampling distribution of $\lambda(\bX_\gamma)$ and directly find the $p$ value for the RetII observation (the fraction of fake data sets with $\lambda(\bX_\gamma)$ larger than the RetII value). This exact $p$ value is compared with the approximate $p$ value obtained using a $\chi^2$ distribution with 2 degrees of freedom. In these experiments we find the actual $p$ to fall between 0.4 and 0.8 times the approximate $p$ value, indicating that the contours are conservative (i.e. the probability they enclose the true value of $\alpha$ and $\beta$ is greater than 68\% and 95\%). In terms of ``sigma values'' (where $1\sigma=68.3\%$ and $2\sigma=95.4\%$), the contours correspond to sigma values about $0.1\sigma$ to $0.3\sigma$ higher than stated. This is perhaps expected since the $\chi^2$ approximation should break down when the true parameters are at the boundary of the parameter space (or beyond).

Figure~\ref{fig:logparabola3fgl} also shows the spectral parameters of 298 sources which are assigned a log parabola spectrum in the 3FGL catalog. Different markers denote different source classes, as described in Table 6 of~\cite{2015ApJS..218...23A}. Extragalactic and unassociated sources are listed in the legend on the left and galactic sources on right.

Of the 395 curved 3FGL sources 114 have one or more analysis flags set, indicating that some aspect of their analysis is problematic (e.g. detection significance or measured flux unstable to changes in the diffuse model, located near a brighter source, poor quality of spectral fit; see~\citep{2015ApJS..218...23A} for details). The majority of these are unassociated sources, have $\beta > 0.25$, and are located very close to the Galactic plane where the source density is high and the diffuse model more uncertain. We remove the 97 sources with $|b|<5\degr$ that have an analysis flag set (other than the flag indicating $\beta=1$). The remaining 17 flagged sources are shown with faded markers in Fig.~\ref{fig:logparabola3fgl}. This selection removes sources with likely biased parameters that are anyway unlikely to be counterparts of a source at RetII's location ($b \approx -50\degr$).

Error bars on the individual 3FGL sources are omitted for clarity but we note that the sizes of the errors on $\alpha$ and $\beta$ are each highly correlated with the value of $\beta$. For the sources in each of four $\beta$ bins ($\beta \in [0, 0.25]$, $[0.25,0.5]$, $[0.5,0.75]$, $[0.75,1]$) we show the median uncertainty on $\alpha$ and $\beta$ as a series of error bars running up the right-hand side of the figure.

The RetII contours are quite large compared to the 3FGL error bars because RetII is detected at much lower significance than these sources. There are a number of unassociated 3FGL sources (empty black circles) within the RetII $2\sigma$ Pass 7 contour, and even several BL Lacs (empty blue diamonds) and active galaxies of uncertain type (filled green circles). The vast majority of sources that can be associated with galactic or extragalactic counterparts, however, have significantly different spectral shape than RetII. In particular, the two blazar classes (BL Lacs and flat spectrum radio quasars) that make up the bulk of associated curved sources populate a relatively well-defined region in $\alpha, \beta$ space with $\beta \lesssim 0.3$. Of the 2918 non-pulsar 3FGL sources, 298 (with $|b|>5\degr$ and no analysis flag) have significant curvature and of these only 5 (40) lie within the 68\% (95\%) Pass 7 RetII contours. Even with the limited photon counts, the data suggest that RetII may have spectral parameters substantially different from almost all other known gamma-ray sources.

\begin{figure*}
\includegraphics{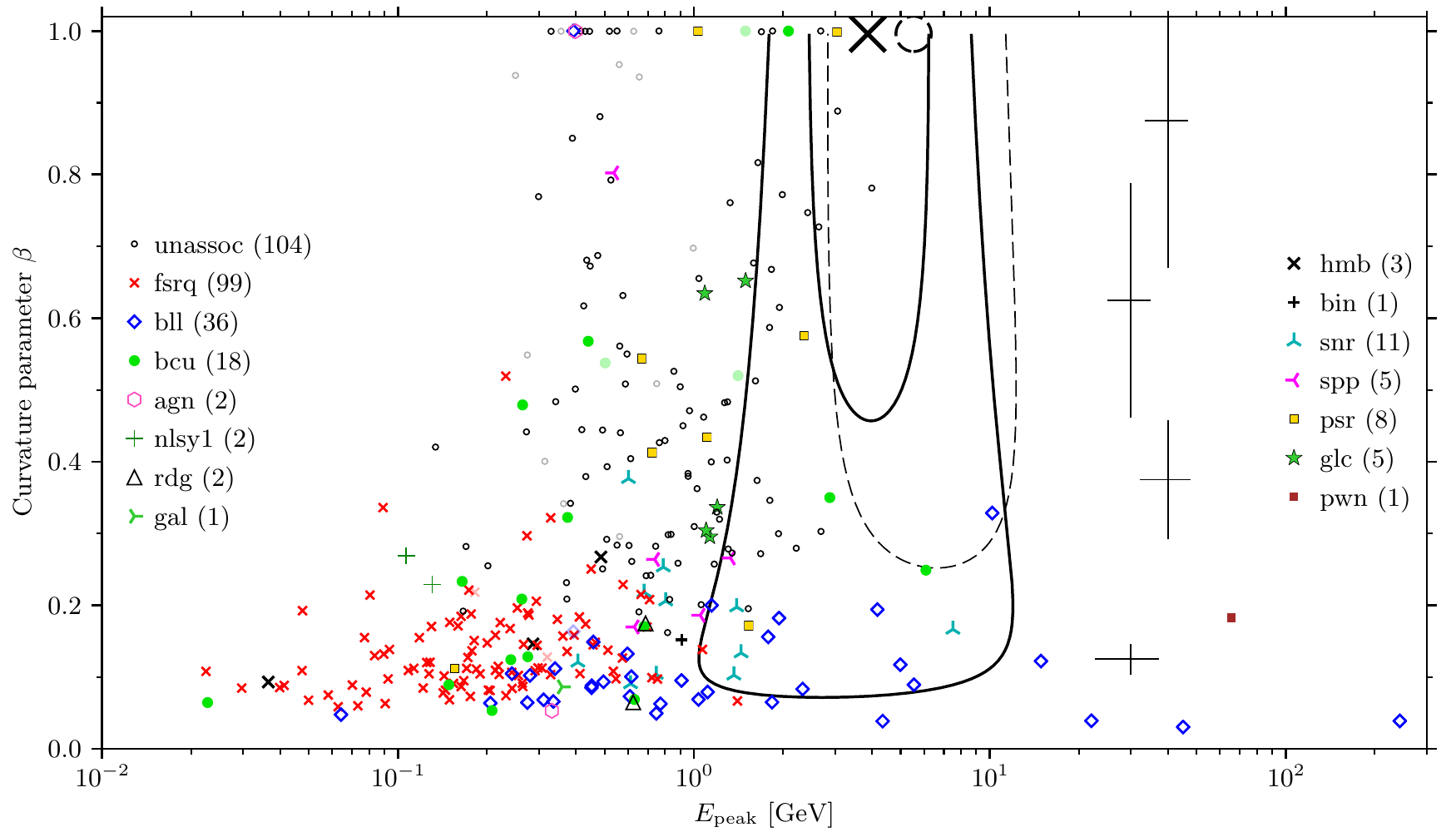}
\caption{\label{fig:logparabola3fgl_Epeakbeta} Same as Fig.~\ref{fig:logparabola3fgl} but with log parabola spectra parameterized by $\Epeak$ (the energy at which the energy flux $E^2 dF/dE$ peaks) and curvature $\beta$.}
\end{figure*}

From Eq.~\ref{eqn:logparaboladef} we see that the maximum of $E^2 dF/dE$ (the spectral energy distribution) occurs at an energy $\Epeak$ where $\log(\Epeak/E_0) = (1 - \alpha/2)/\beta$. In Fig.~\ref{fig:logparabola3fgl} contours of constant $\Epeak$ (gray dashed lines) are straight lines radiating from $(\alpha,\beta)=(2,0)$, with positive slope if $\Epeak < E_0$ and negative slope if $\Epeak > E_0$. The degeneracy direction in the $\alpha, \beta$ contours suggests that $\Epeak$ is what is actually being measured in the data, rather than $\alpha$ and $\beta$ individually. This is verified in Fig.~\ref{fig:logparabola3fgl_Epeakbeta}, where we reparameterize the log parabola spectra using $\Epeak$ instead of $\alpha$ and find the best fit within the range $10~\MeV < \Epeak < 1~\TeV$ and $0<\beta<1$ . Representative error bars for 3FGL sources are obtained with the same binning procedure used in Fig.~\ref{fig:logparabola3fgl} (we use simple error propagation to find the errors on $\log(\Epeak/E_0)$ and note that for some sources with the lowest measured $\beta$'s the errors on $\Epeak$ can reach 100\%, reflecting the fact that $\Epeak$ is ill-defined as $\beta\rightarrow 0$). While the data imply a lower limit on RetII's curvature parameter, they provide a well-constrained measurement of the peak of its spectral energy distribution as expected from, e.g., Fig.~\ref{fig:speckstackmodels}. It appears that FSRQs radiate most of their gamma-ray energy in photons of systematically lower energy than RetII does.

It is clear from Fig.~\ref{fig:logparabola3fgl_Epeakbeta} that curvatures larger than $\beta=1$ will improve the fit. If we relax the constraint from the 3FGL that $\beta \leq 1$, the best fitting $\beta$ increases from 1 to 3.0 in Pass 7 (3.8 in Pass 8), and $\lambda(\bX_\gamma)$ becomes similar to the best fitting models of Table~\ref{tab:gof78}. Since $\Epeak$ and $\beta$ are uncorrelated we measure them each individually, maximizing the likelihood over the other parameter and $F_0$, and assuming that the likelihood ratio is governed by a $\chi^2$ distribution with 1 degree of freedom. We measure $3.1 < \Epeak /\GeV < 4.5$ and $1.4 < \beta < 5.9$ at 68.3\% confidence for the Pass 7 data. For Pass 8 we find $4.0 < \Epeak /\GeV < 6.4$ and $\beta > 1.6$. As with Pass 7, the likelihood ratio rises rapidly as $\beta$ decreases from its best fit value, but $\lambda(\bX_\gamma)$ is still less than 1 when $\beta=10$, yielding a one-sided confidence interval on $\beta$. As expected, $\Epeak$ is constrained rather precisely while the data essentially provide a lower bound on $\beta$.

\subsection{Pulsars  \label{sec:pulsars}}

Among the source classes in the 3FGL, pulsars are notable for their significantly curved spectra. About 75\% of the pulsars in the 3FGL have a curvature significance greater than $4\sigma$ (as compared with FSRQs (17\%), BL Lacs (3\%), blazars of uncertain type (3\%), supernova remnants (57\%), globular clusters (40\%), and unassociated sources (17\%)). If RetII hosts one or more gamma-ray emitting pulsars that may explain its curved spectrum.

We make an estimate of the pulsar contribution to RetII's gamma-ray flux by considering the 15 globular clusters in the 3FGL. The gamma-ray emission from globular clusters is likely powered by populations of millisecond pulsars (MSPs)~\cite[e.g.][]{1991Natur.352..695C,2007MNRAS.377..920B,2010A&A...524A..75A, 2011Sci...334.1107F,2013ApJ...778..106J}. For each globular cluster, we scale its gamma-ray flux to what it would be at the distance of RetII. We also scale the gamma-ray flux according to the ratio of V-band luminosity $L_V$ of the globular cluster~\citep[][2010 edition]{1996AJ....112.1487H} to that of RetII~\citep{2015ApJ...805..130K}. In this way each globular cluster provides an estimate of the pulsar emission which might be expected from RetII. As RetII has an old, metal-poor stellar population, similar to globular clusters, the visual luminosity is a proxy for number of stars. The luminosity scaling assumes that the number of MSPs in a system is proportional to the number of stars. However, MSPs are typically found in binary systems, and the stellar encounter rate in globular clusters correlates with both their abundance of X-ray binaries (possible progenitors of MSPs) and their gamma-ray luminosity~\cite{2016JASS...33....1T,2010A&A...524A..75A,2013ApJ...766..136B}.  Because of its extremely low stellar density compared to globular clusters (hence low encounter rate), RetII likely harbors a far smaller fraction of binary systems than do globular clusters, and therefore a far fewer number of MSPs per unit luminosity. Furthermore, selection bias leads to the globular clusters in the 3FGL having higher gamma-ray luminosities than expected from the scaling based on the population of Milky Way globular clusters (there are 104 gamma-ray quiet globular clusters in the Harris catalog with greater $L_V/D^2$ than the 3FGL globular cluster with the smallest $L_V/D^2$). For these reasons, our globular cluster scaling is conservative and will tend to overestimate the RetII pulsar flux.

\begin{figure}
\includegraphics{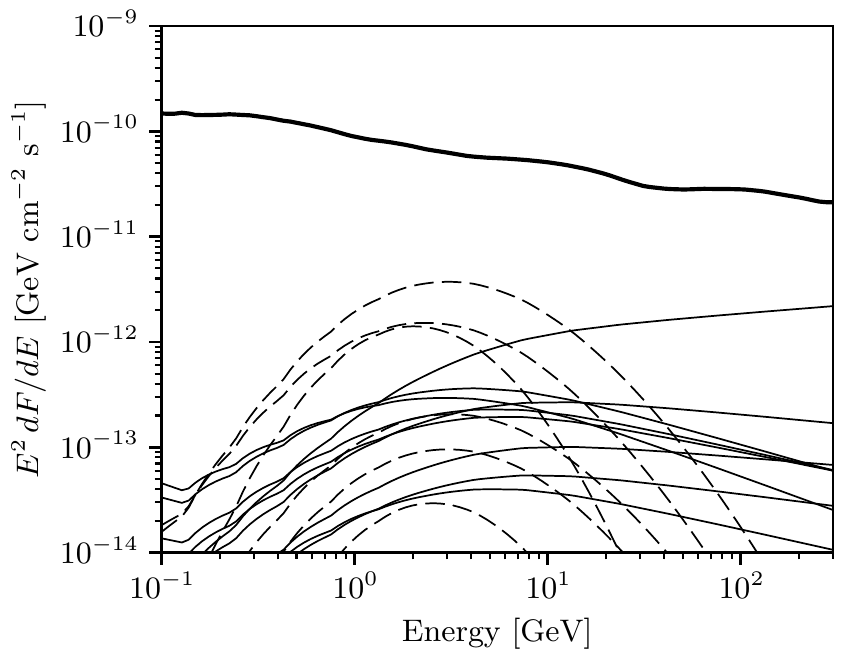}
\caption{\label{fig:GCs} The pulsar contribution to RetII's gamma-ray flux compared with the diffuse background level. Each thin curve shows the spectrum of a gamma-ray detected globular cluster scaled according to the distance and luminosity of the globular cluster relative to RetII. Solid lines correspond to globular clusters with power law spectra in the 3FGL, dashed lines to those with log parabola spectra. The thick curve shows the diffuse background. Fluxes are integrated over a region of radius $0.25\degr$ as in Fig.~\ref{fig:speckstackmodels}.}
\end{figure}

Figure~\ref{fig:GCs} compares the diffuse background level toward RetII to the estimates of pulsar emission provided by each scaled globular cluster. Fluxes are shown integrated within $0.25\degr$ of RetII (compare with Figs.~\ref{fig:speckstackmodels} and~\ref{fig:speckstackannuli}). While the spectral shape of globular cluster emission is often quite similar to what we observe from RetII, the expected flux is too small to explain the RetII signal by over an order of magnitude. We also consider the peak intensity (flux per solid angle) of the scaled globular cluster emission: the maximum value of the PSF multiplied by the point source flux (see Eq.~\ref{eqn:signalfluxpoint}). Except for Palomar~9, each scaled globular cluster has a gamma-ray intensity an order of magnitude or more below the background estimate in RetII's direction. Palomar 9's intensity lies slightly above background at energies above 30 GeV. However, at these energies we expect fewer than a single event to be detected by Fermi. We conclude that it is highly unlikely that a population of MSPs could give rise to an observable gamma-ray signal from RetII.

Another way to see the implausibility of the MSP explanation is to note that the estimated number of MSPs in gamma-ray emitting globular clusters range from about ten to at most a few hundred~\cite{2010A&A...524A..75A,2016MNRAS.459...99Z,pulsarsinGCswebsite}. This relative handful of MSPs occur in densely packed systems of millions of stars. RetII, with about 1000 solar luminosities, is unlikely to possess a single MSP. In fact, using a sample of globular clusters {\em not} selected by gamma-ray luminosity,~\citet{2016JCAP...08..018H} find the occurrence of MSPs in globular clusters to be about 1 per $10^6$ solar luminosities. 

The results of this section, based on simple scaling arguments, are in agreement with the conclusions of~\citet{2016ApJ...832L...6W}. In that study, the pulsar contribution to dwarf galaxy gamma-ray fluxes is estimated by constructing a gamma-ray luminosity function for isolated Milky Way MSPs and then scaling the Milky Way population down by the ratio of dwarf galaxy to Milky Way stellar mass. The authors find an expected pulsar contribution 1 to 5 orders of magnitude below the diffuse background gamma-ray flux for all ultrafaint dwarf galaxies.

Our arguments do not address the possibility of an MSP unrelated to RetII that happens to lie along the line of sight. The probability of such a coincidence can be estimated with population synthesis simulations~\cite[e.g.][]{2010JCAP...01..005F,2011MNRAS.415.1074S,2013A&A...554A..62G,2014ApJ...796...14C,2016ApJ...832L...6W}. In particular, \citet{2016ApJ...832L...6W} predict the flux from foreground pulsars to be similar to the flux from those internal to RetII. Dedicated searches in radio~\cite[e.g.][]{2017JCAP...07..025R} and X-rays may also be able to discover an interloping pulsar.

\section{Discussion \label{sec:discussion}}

Among the types of background gamma-ray sources that might lie along RetII's line of sight, blazars are perhaps the most likely candidates. This population makes up the majority of associated 3FGL sources at high galactic latitudes. In addition, a large fraction of the unassociated 3FGL sources likely have blazar counterparts~\citep[e.g.][]{2012ApJ...752...61M,2012ApJ...753...83A,2016PhRvL.116o1105A,2015PhR...598....1F}. Generic radio sources~\citep{2003MNRAS.342.1117M,2012MNRAS.422.1527M} number in the hundreds of thousands, with only a tiny fraction being associated with a gamma-ray source. In contrast, around 30\% of the approximately 3000 known blazar candidates appear in the 3FGL~\citep{2009A&A...495..691M,2015ApJ...810...14A}. As we have shown, comparing the gamma-ray spectra of blazars with that of a tentative gamma-ray source offers a way of making a distinction between dark matter annihilation and blazar emission. Taken at face value, the high curvature and spectral energy peak of RetII are markedly different from the two main blazar types. In particular, $\Epeak$ is measured more robustly than $\beta$ making the separation between RetII and the FSRQs especially clean. The separation between RetII and the BL Lacs is based on the apparently large curvature of RetII, though this separation is less marked than for the FSRQs. This is particularly important as \citet{2017JCAP...07..025R} have identified two BL Lac candidates behind RetII. As Fermi increases its exposure (and if there is in fact a source in RetII's direction), confidence regions in Fig.~\ref{fig:logparabola3fgl_Epeakbeta} will shrink, the catalog of gamma-ray loud blazars will expand, and the comparison between RetII and blazar types will come into sharper focus.

In contrast to blazars, our analysis of gamma-ray spectral shape cannot distinguish a dark matter signal from pulsar emission. This spectral similarity has been a central issue in the search for dark matter annihilation at the Galactic Center~\citep[e.g.][]{2011JCAP...03..010A,2013PhRvD..88h3009H,2016PhRvL.116e1102B,2016PhRvL.116e1103L}. Searches in dwarf galaxies appear to avoid the problem (Sec.~\ref{sec:pulsars}, also~\citep{2016ApJ...832L...6W}). Of course, RetII may be a system with a peculiar history~\cite[e.g.][]{2016ApJ...830...93J,2017arXiv170706871S} and our analogy with globular clusters may break down. Future study of RetII at all wavelengths will help pin down possible gamma-ray sources within (and behind) RetII.

Our discussion in Sec.~\ref{sec:significance} about the relative probabilities of a statistical fluctuation vs. an additional gamma-ray source in RetII's direction is based on a well known property of the Fermi sky: sampling random sky locations turns up more ``high-significance'' locations (i.e. hot pixels) than would be expected from the Poisson statistical fluctuations of the diffuse model (see, e.g.~\citep{2015PhRvD..91h3535G,2014PhRvD..89d2001A,2015PhRvD..91f1302C,2015PhRvL.115w1301A,2017ApJ...834..110A} in the context of dark matter searches). This phenomenon has been invoked to argue for a millisecond pulsar explanation of the Galactic Center gamma-ray excess~\citep{2015JCAP...05..056L,2016PhRvL.116e1102B,2016PhRvL.116e1103L} and, at higher latitudes, to constrain source populations~\citep{2010JCAP...01..005F,2011ApJ...738..181M,2014ApJ...796...14C,2016ApJ...826L..31Z,2016ApJS..225...18Z,2016PhRvL.116o1105A,2017ApJ...839....4D,2016ApJ...832..117L} as well as dark matter annihilation~\citep{2009JCAP...07..007L,2010PhRvD..82l3511B,2015JCAP...09..027F,2017arXiv171001506Z}. In this work, the task is to understand the origin of one particular hot pixel (i.e. RetII) that is known to host a dark matter halo with a large $J$ value. As the origin of the excess high-significance locations becomes better understood (i.e. via ``1-point function'' analyses) it will be possible to quantify the probability that a RetII-like observation is caused by a particular class of sources.

Finally, we return to the differences between Pass 7 and Pass 8. There are differences in detection significance and in the best fitting properties of the RetII source when analyzed with the two data sets. The ultimate solution requires finding the probability of jointly obtaining the Pass 7 and Pass 8 results when there either is or is not a source in RetII's direction. This is beyond the scope of this work. However, our analysis can partially address the consistency question: is there a single RetII energy spectrum consistent with both the Pass 7 and Pass 8 data sets?

Though we analyzed the two data sets independently, in reality they are highly overlapping~\cite[e.g.][]{2015PhRvL.115w1301A}: dividing the events detected within $15\degr$ of RetII into energy bins we find that, over the same 6.9 observation, about 60-70\% of events between 1 and 10~GeV found in one data set are also found in the other. 

We can obtain a necessary condition for consistency by treating the Pass 7 and Pass 8 data sets as two independent observations of the same object. If there is no single spectrum that is a good fit to both data sets when they are treated independently then the data sets will certainly be inconsistent had their dependence been properly included. The best fitting log parabola model to the combined Pass 7/Pass 8 data set (with all three log parabola parameters completely free) is $\alpha=-8.1$, $\beta=3.5$, $\Epeak=4.2\,\GeV$, and $F_0 = 1.5\times 10^{-13} \,\cm^{-2} \sec^{-1} \GeV^{-1}$. Taking this model as the null model $\theta_0$ (Sec.~\ref{sec:GOFmethod}) the goodness of fit test statistic is $\lambda(\bX_\gamma)=1.1$ for the Pass 7 data, and $\lambda(\bX_\gamma)=1.0$ for the Pass 8 data. The distribution for $\lambda(\bX_\gamma)$ should be $\chi^2$ distributed with 3 degrees of freedom, giving $p\approx 0.8$ for both Pass 7 and Pass 8. When considered independently, the confidence intervals for the log parabola parameters inferred from Pass 7 and 8 are therefore highly overlapping. Correlations between the two data sets will increase the level tension but we conclude that consistency is plausible.

There are also indications of consistency between 6.9 years of Pass 7 the 9-year Pass 8 results of \citet{2018arXiv180506612L}. The best fitting dark model reported by~\cite{2018arXiv180506612L} is for dark matter with a mass of 16~GeV annihilating into $\tau$ leptons. While they do not report a best fitting annihilation cross section, this mass fits the 6.9-year data essentially as well as the best fitting masses we find ($M=13.9~\GeV$ in Pass 7 and 19.5~GeV in 6.9 years of Pass 8; $\lambda(\bX_\gamma)$ changes by $\sim 0.1$ when shifting these masses to 16~GeV while maximizing with respect to $\sigv J$). As for level of significance, the reported $TS=13.5$ corresponds to $p \approx 10^{-4}$ for the diffuse Poisson background model as mentioned in the introduction. Using Fig.~11 of~\citep{2017ApJ...834..110A}, which is based on a blank sky calibration that takes into account a trials factor needed for searching multiple masses, such a $TS$ value corresponds to $p\approx 0.01$. This blank sky method for evaluating significance is approximately analogous to the empirical background sampling method which yielded $p=0.01$ in 6.5 years of Pass 7 data~\cite{2015PhRvL.115h1101G}. Thus the arguments of Sec.~\ref{sec:significance} may well hold for the 9-year Pass 8 data as well as the 6.5-year Pass 7 data.

\section{Conclusions}

We present a series of analyses to follow up on the detection of a gamma-ray signal from the direction of a dwarf galaxy. Our main focus is on assessing whether there is a plausible astrophysical interpretation for the signal. We first quantify the probability that the excess is due to a Poisson fluctuation of diffuse background processes vs. the existence of a previously unknown point source source along the line of sight. We show that comparing the gamma-ray spectrum of the new source to those of known classes of gamma-ray emitters can help rule out a chance alignment with an unrelated background object. Finally, we estimate the level of emission from a population of pulsars within the dwarf which could mimic a dark matter signal.

These analyses are applied to Fermi observations of the Reticulum~II dwarf, the most promising dwarf dark matter signal seen so far. We find that a line of sight featuring a gamma-ray excess like RetII's has high likelihood (probability greater than 99\%) of hosting a gamma-ray source with flux above the diffuse background level. We use a simple log parabola parameterization of RetII's gamma-ray spectrum and compare with known sources in the 3FGL catalog. RetII has a significantly curved energy spectrum, which is a distinctive feature among gamma-ray sources. We find that of the blazar types (which represent the majority of high latitude associated gamma-ray sources), flat spectrum radio quasars emit most of their gamma-ray energy at lower energy ($\Epeak$) than RetII does. BL Lacs can emit at energies comparable to RetII's, though they in general have spectral curvatures too low to explain the RetII data. All of these conclusions are stronger when considering 6.9 years of Pass 7 data than the same amount of Pass 8 data, for which the significance of the RetII excess is lower and all confidence regions expand considerably. For any promising dark matter target, not just RetII, these techniques will help to distinguish a dark matter explanation from an astrophysical one.

\begin{acknowledgments}
We acknowledge useful discussions with Kev Abazajian, Gordon Blackadder, Ian Dell'Antonio, Raphael Flauger, Sebastien Fromenteau, Rick Gaitskell, Manoj Kaplinghat, Fran\c{c}ois Lanusse, Sandhya Rao, Pat Scott, Sukhdeep Singh, Louie Strigari, Roberto Trotta, and Aaron Vincent. SMK is supported by DE-SC0017993. MGW is supported by NSF grants AST-1313045 and AST-1412999. .

\end{acknowledgments}

\bibliography{bibfile}

\end{document}